\documentclass[11pt]{article}
\usepackage{amsmath,amssymb,amsfonts,color,epsf}
\usepackage{amssymb,slashed,latexsym,multirow,color}
\usepackage{graphics}
\makeatletter
\@addtoreset{equation}{section}
\makeatother

\newcommand{\hoch}[1]{$\, ^{#1}$}

\newcommand{\be}{\begin{equation}}
\newcommand{\ee}{\end{equation}}
\newcommand{\bea}{\setlength\arraycolsep{2pt} \begin{eqnarray}}
\newcommand{\eea}{\end{eqnarray}}
\newcommand{\nn}{\nonumber}

\usepackage[font=footnotesize,labelsep=newline,labelfont=sc,justification=centering,position=top]{caption}

\usepackage{hyperref}

\newsavebox{\uuunit}
\sbox{\uuunit}
    {\setlength{\unitlength}{0.825em}
     \begin{picture}(0.6,0.7)
        \thinlines
        \put(0,0){\line(1,0){0.5}}
        \put(0.15,0){\line(0,1){0.7}}
        \put(0.35,0){\line(0,1){0.8}}
       \multiput(0.3,0.8)(-0.04,-0.02){12}{\rule{0.5pt}{0.5pt}}
     \end {picture}}

\def\be{\begin{equation}}
\def\ee{\end{equation}}
\def\ba{\begin{array}}
\def\ea{\end{array}}
\def\bea{\begin{eqnarray}}
\def\eea{\end{eqnarray}}
\def\bd{\begin{displaymath}}
\def\ed{\end{displaymath}}

\def\nn{\nonumber}

\def\t{\tau}

\def\x{\xi}

\def\nn{\nonumber}

\begin{document}

\begin{flushright}
\hfill{ \
\ \ \ \ UG-2014-31 \ \ \ \ \\
\ \ \ \ ICCUB-14-050  \ \ \ }
\end{flushright}
\vskip 1.2cm

\begin{center}
{\Large \bf
{
Dynamics of Carroll Particles}}
\end{center}
\vspace{25pt}
\begin{center}
{\Large {\bf }}

\vspace{10pt}

{\Large Eric Bergshoeff\hoch{1}\,,
 Joaquim Gomis\hoch{2} and Giorgio Longhi\hoch{3}
 }

\vspace{10pt}

\hoch{1} {\it Centre for Theoretical Physics, University of Groningen,\\
Nijenborgh 4, 9747 AG Groningen, The Netherlands}\\

 \hoch{2} {\it Departament d'Estructura i Constituents de la Mat\`eria and Institut de Ci\`encies del\\
Cosmos, Universitat de Barcelona, Diagonal 645, 08028 Barcelona,
Spain}

\hoch{3} {\it Department of Physics and Astronomy, University of Florence, 50019 Sesto F., Firenze,\\
Italy;Sezione di Firenze, INFN, 50019 Sesto F., Firenze, Italy}

\vspace{30pt}

\underline{ABSTRACT}
\end{center}

We investigate particles whose dynamics is invariant under the Carroll group. Although a single free such Carroll particle has no non-trivial dynamics (`the Carroll particle does not move') we show that there exists non-trivial dynamics
for a set of interacting Carroll particles. Furthermore, we gauge the Carroll algebra and couple the Carroll
particle to these gauge fields. It  turns out that for such a coupled system even a single Carroll particle can have non-trivial dynamics.

\vspace{15pt}

\thispagestyle{empty}

\vspace{15pt}

 \vfill

\thispagestyle{empty}
\voffset=-40pt

\newpage

\tableofcontents


\newpage


\section{Introduction}

A long time ago Bacry and Levy-Leblond \cite{Bacry:1968zf} have classified the possible relativity groups in four dimensions.
A relativity group is here defined as a possible invariance group of a 4D physical theory that contains the generators of relativity, i.e.~time translations, space translations, spatial rotations and boosts.  Apart from the well-known Poincar\'e and Galilei groups their classification also yielded the  adS, dS and Newton-Hooke groups \cite{dubouis} and, furthermore, the less well-known  Carroll groups \cite{Levy-Leblond}\footnote{In that time supersymmetry was not yet discovered.}. They also proved  how these different relativity groups are related by  In\"on\"u-Wigner contractions. In fact, all of them can be obtained by a contraction of the adS and dS groups. It turns out that there are  three different types of contractions of these groups: (1) the one that takes the radius of curvature of the adS/dS spacetime to infinity; (2)  the non-relativistic contraction that takes the velocity of light to infinity and (3)  the Carroll contraction that takes the velocity of light to zero. In some sense the Carroll contraction is the opposite of a non-relativistic contraction and can be viewed as the ultra-relativistic limit. Such ultra-relativistic limits have been studied in, e.g., \cite{Henneaux:1979vn,Dautcourt:1997hb}. Recently, it has been pointed out that there is an interesting relationship between  the Carroll symmetries and the BMS algebra \cite{Duval:2014uva}.
The BMS algebra \cite{Bondi:1962px}
has emerged  recently as the boundary symmetry group  in a study of flat space holography,
see for example
\cite{Barnich:2009se,Bagchi:2012cy}.

It is the purpose of this paper to study the dynamics of particles that realize the Carroll symmetries. It is not obvious  that such `Carroll particles'  allow for any non-trivial particle dynamics.
In fact, it turns out that the free Carroll particle  has no dynamics at all: the free Carroll particle  does not move  \cite{old,  Ngendakumana:2013hza, Duval:2014uoa}.
 In this paper we wish to investigate whether this is the generic situation or whether Carroll symmetries can allow for any non-trivial particle dynamics. We will first study the free Carroll particle. In particular, we
will show that  the mass-shell constraint allows for positive and negative energies and that  the energy is proportional to the mass of the particle, it can be positive, negative or zero. The latter case    corresponds to massless Carroll particles. The quantization of the free Carroll particle leads to a kind  of ultra-local relativistic Poincar\'e theory.
 We will also study  the symmetries of free Carroll particles and conclude that the Carroll symmetry is enlarged to an infinite dimensional symmetry.
 The particle models we consider in this paper are obtained by taking the Carroll limit of the relativistic particle. As we will see,
this Carroll limit wipes out all information about the curvature of the spacetime in which the original relativistic particle was moving in.

 Next, we will extend the analysis to  two-particle systems in a flat spacetime and show that, in contrast to the single particle case, there is non-trivial dynamics. Moreover, we will show that the infinite-dimensional
  symmetry of the free Caroll particle collapses to a finite-dimensional global Carroll symmetry.
We will describe one more situation in which the Carroll particle can have non-trivial dynamics.
That happens when we gauge the Carroll algebra and use this as input to construct the  coupling of the Carroll gauge fields to the particle. We will show that the Carroll particle in such a non-trivial background can have non-trivial dynamics.

The organization of this paper is as follows. In section 2 we introduce the free Carroll particle and identify its infinite dimensional symmetries. In section 3 we extend the analysis and consider a model of two interacting Carroll particles.
In the next section we investigate the dynamics of this model and show that, unlike the free case,  there is non-trivial dynamics.
Subsequently, in section 5 we gauge the Carroll algebra and introduce the Carroll gauge fields. In the same section
we consider the
coupling of these gauge fields to the Carroll particle  and again we will show that there is non-trivial dynamics.
 Finally, the conclusions are given in section 6.



\section{The Free Carroll Particle}

One way to obtain the action of the free Carroll particle is to start  from the massive particle moving  in an adS or dS spacetime and to take the Carroll limit\,\footnote{
The action for the free Carroll particle can alternatively  be obtained by using the method of non-linear realizations \cite{Coleman} applied to the Carroll algebra \cite{old} or by applying the method of  coadjoint orbits \cite{Duval:2014uoa}. More details about the first construction can be found in the appendix.}.
The canonical form of the action before taking the limit is given by\,\footnote{  The signature of the metric is $(-,+,+,+).$}
\be\label{ads}
S=\int d\tau [ p\cdot \dot x-\frac {e}{2}(p^2+m^2)]\,,
\ee
\color{black}
where $p^2=g^{\mu\nu}(x)p_{\mu} p_\nu$ and $g^{\mu\nu}(x)$ is the inverse metric of an adS or dS space. We will work in a basis in which the  adS
 line element is given by
\bea\label{AdS}
ds^2&=&- \cosh^2{\frac{r}{R}}(dX^0)^2+\left(
\frac{\sinh{\frac{r}{R}}}{{\frac{r}{R}}}\right)^2 (dX^a)^2-
\left(\left( \frac{\sinh{\frac{r}{R}}}{{\frac{r}{R}}}\right)^2-1\right)
(dr)^2\,,
\eea
where $r=\sqrt{X_a X^a}$. Similarly, the dS line element is given by
\bea\label{dS}
ds^2&=&+ \cos^2{\frac{r}{R}}(dX^0)^2+\left(
\frac{\sin{\frac{r}{R}}}{{\frac{r}{R}}}\right)^2 (dX^a)^2-
\left(\left( \frac{\sin{\frac{r}{R}}}{{\frac{r}{R}}}\right)^2-1\right)
(dr)^2.
\eea

We next consider the  Carroll limit which  is given by\,\footnote{Note that we use a dimensionless parameter $\omega$  instead of the velocity of light. Indeed, if one considers the Carrollian counterpart of the non-relativistic limit of a string,  one needs to use a dimensionless parameter \cite{Gomis:2000bd,Gomis:2005pg}.}\footnote{The contractions we consider in this work correspond to the ultra-relativistic limit of a world probed by particles. There are more general contractions possible that correspond to the ultra-relativistic limit of extended objects such as strings and branes.}
\be
x^0= \frac{t}{\omega}, \quad p^0=\omega E, \quad m=\omega M
\ee
with $\omega\to\infty$. It is understood that, before taking the limit,  we rescale the Einbein variable like
\bea
e\to\frac {-e}{\omega^2}
\eea
 for both the AdS and dS cases. The Carroll limit is in both cases the same and is given by
\begin{eqnarray}\label{0}
S_{C} &=&  \int d\tau\, \big[ - E \dot{t}
+ \dot{\vec{x}}\cdot\vec{p}  -\frac{e}{2} (E^2-M^2)\big]\,.
\end{eqnarray}
The canonical action \eqref{0} is invariant under the Carroll transformations
\bea\label{Ctransf'}
&&t'=t+\vec\beta\cdot R\vec x+a_t\,,\hskip 2truecm
\vec x'=R\vec x+\vec a\,,
\nn\\
&&\vec p'=R\vec p+ \vec\beta E\,, \hskip 2.9truecm
E'=E\,.
\eea

We observe  that the Carroll limit has eliminated the $R$-dependence of the relativistic particle action and, consequently, any sign of the curved space time  we started with.
 This implies that the curvature of the transverse space cannot be probed by the Carroll particle. The equations of motion corresponding to the action \eqref{0}  are rather trivial:  the
free Carroll particle is at rest and does not move.

The situation is rather different if we consider instead the  non-relativistic limit  of a particle moving in an AdS or dS spacetime.  In the AdS case, taking the non-relativistic limit leads to the harmonic oscillator with frequency
$\omega=1/R, $ whereas  in the dS case we obtain the inverse harmonic oscillator.
To be specific,  upon performing the rescaling
\be
x^0= {\omega}t, \quad p^0=\frac{E}{\omega}, \quad m=\omega M
\ee
and taking the limit $\omega\to\infty$ in  the canonical action \eqref{ads} with the AdS line-element \eqref{AdS} we obtain\,\footnote{Following \cite{Gomis:2000bd},  in order to eliminate a divergent piece  we introduce the coupling to a constant electromagnetic field.}

\begin{eqnarray}
S_{C} &=&  \int d\tau( - E \dot{t} + \dot{\vec{x}}\cdot\vec{p}  -e (2ME-\vec p^2-\frac{1}{R^2}\vec x^2).
\end{eqnarray}
Taking the same limit in the action \eqref{ads} with the de Sitter line element \eqref{dS} we obtain  the following action:
\begin{eqnarray}\label{dsnr}
S_{C} &=&  \int d\tau( - E \dot{t}
+ \dot{\vec{x}}\cdot\vec{p}  -e (2ME-\vec p^2+\frac{1}{R^2}\vec x^2),
\end{eqnarray}
which is the inverse harmonic oscillator.
Therefore, in contrast to the Carroll limit, the relativistic limit keeps track of
the information of the curved space we started with.

Going back to the Carroll particle, the mass-shell constraint of the free Carroll particle  is given by
\be
\phi=E^2-M^2=0\,,\label{mass-shell}
\ee
which is solved by $E=\pm M$.
Note that the mass-shell constraints do not depend  on the spatial momenta.
A difference with respect to the non-relativistic case is that here we can have particles with negative energy
and, furthermore, the  massless limit $M \rightarrow 0$ is well-defined.  The action of a massless Carroll particle becomes
\begin{eqnarray}\label{00}
S_{C} &=&  \int d\tau( - E \dot{t}
+ \dot{\vec{x}}\cdot\vec{p}  - \frac{e}{2} E^2)\,.
\end{eqnarray}

If we quantize the model and impose the mass-shell constraint on the physical states we end up with a kind of  ultra-local relativistic Poincar\'e theory
with wave equation
\be
\Big(-\frac {d^2}{dt^2}-M^2\Big)\phi(t,\vec x)=0\,.
\ee

Solving for the mass-shell constraints \eqref{mass-shell} we can write the action in the equivalent form
\begin{eqnarray}\label{1}
S_{C} &=&  \int d\tau\, \big[ \mp M\dot{t}
+ \dot{\vec{x}}\cdot\vec{p}\big]\,.
\end{eqnarray}
In this case the transformation of the momenta becomes $\vec p'=R\color{black}\vec p\pm \vec\beta M$.

The generators of the Carroll algebra are given by
\be
H=E\,,\hskip 1truecm \vec P = \vec p\,,\hskip 1truecm \vec G = E\vec x\,,\hskip 1truecm \vec J = \vec x \times \vec p\,,
\ee
which is to be supplemented by the mass-shell condition $E= \pm M$. In the massless case the generators are given by
\be
H=0\,,\hskip 1truecm \vec P = \vec p=p\vec u\,,\hskip 1truecm \vec G = 0\vec x\,,\hskip 1truecm \vec J = \vec x \times \vec p\,,
\ee
where $\vec u$ is a unit vector, i.e.~a general element of $S^2$ \cite{Duval:2014lpa}.

We will now dedicate a separate subsection to a discussion of the symmetries of the free Carroll particle.

\subsection{Infinite-dimensional Symmetries}

The basic Poisson brackets of the canonical variables occurring in the action \eqref{0} are given by
\be
\{E,\ t\} = 1, \quad \{e, \  \pi\} = 1,\quad \{x_i^m,\ p_j^n\} = \delta_{m,n}\delta^{i,j}\,.
\ee
This leads to the following equations of motion for these variables:
\be
\dot t=-eE,\quad \dot{\vec{x}}=0,\quad \dot e =\lambda(\tau), \quad \dot E=0, \quad \dot{\vec{p}}=0\,, \quad\dot\pi=-1/2(E^2-M^2)\,,
\ee
where $\lambda (\tau)$ is an arbitrary function and $\pi$ is the momenta associated to the Einbein
variable e which is constrained by  $\pi=0$.

Consider now the following generator of canonical transformations
\be
G=-E\xi^0(\vec x,t)+p_i\xi^i(\vec x,t)+\gamma(\vec x,t)\pi,
\ee
with parameters $\xi^0(\vec x,t)\,,\xi^i(\vec x,t)$ and $\gamma(\vec x,t)$.
 The transformations generated by this generator  are given by
\bea
\delta t&=&\xi^0(\vec x,t),\quad \delta x^i=\xi^i(\vec x,t), \quad \quad \delta e=\gamma(\vec x,t)\,,
\nn\\[.1truecm]
 \delta  E&=&  -\partial_t \xi^0(\vec x,t) E+\partial_t \xi^i(\vec x,t)p_i+\partial_t \gamma(\vec x,t)\pi\,,
\nn\\[.1truecm]
 \delta p_i&=&\partial_i\xi^0(\vec x,t) E -\partial_i\xi^j(\vec x,t) )p_j-\partial_i\gamma(\vec x,t)\pi\,.
\eea
These transformations are  symmetries of the free Carroll particle, provided that $G$ is a constant of
motion, i.e., $\partial_\tau G=0$. This leads to the following restriction on the parameters:
\bea
&0&=-E (\dot t\partial_t\xi^0(\vec x,t)+\dot{x_j}\partial^j\xi^0(\vec x,t))+
p_i  (\dot t\partial_t\xi^i(\vec x,t)+\dot{x_j}\partial^j\xi^i(\vec x,t))+\dot\pi \gamma(\vec x,t)\nonumber\\[.1truecm]
&& =-eE^2\partial_t\xi^0(\vec x,t)-eEp_i\partial_t\xi^i(\vec x,t)-\frac{1}{2}\gamma(\vec x,t)(E^2-M^2)\,.
\eea
From this equation we deduce the following Killing equations corresponding to the free Carroll particle:
\be
\partial_t\xi^0(\vec x,t)=0,\quad \partial_t\xi^i(\vec x,t)=0,\quad \gamma(\vec x,t)=0\,.
\ee
The solutions of these Killing equations are
\be\label{solKil}
\xi^0=\xi^0(\vec x),\quad \xi^i=\xi^i(\vec x)
\ee
and, hence, the generator G is given by
\be
G=-E\xi^0(\vec x)+p_i\xi^i(\vec x)\,.
\ee
We thus conclude that the free Carroll particle has an infinite dimensional symmetry.
The  Carroll transformations \eqref{Ctransf'} are obtained  by keeping the first term in the powers series
expansion  of  the parameters $\xi^0(\vec x)$ and $ \xi^i(\vec x)$ in terms of $\vec x$. Adding some curvature structure by hand to the transverse space will  eliminate the "transverse spatial" transformations as in \cite{Duval:2014uoa, Duval:2014uva}.

In the special case of a massless Carroll particle the isometries should be given by
 the most general conformal Carroll group.
The Killing equations in  this case become
\be
\partial_t \xi^0(t,\vec x)-\frac{\gamma}{2e}=0,\quad \partial_t\xi^i(t,\vec x)=0
\ee
for arbitrary parameter $\gamma(\vec x,t)$.
This leads to the following  generator of conformal Killing transformations
\be
G=-E\xi^0(t,\vec x)+p_i\xi^i(\vec x)+2\pi_e \partial_t \xi^0(t,\vec x)\,,
\ee
which again generates an infinite dimensional symmetry.  These transformations include scale transformations of the time and space coordinates. If we put more structure in the transverse space
these transformations will have restrictions   and we could obtain the
Carroll trasformations of
\cite{Duval:2014lpa, Duval:2014uva}.

This concludes our discussion of the free Carroll particle, its dynamics  and its symmetries.
\section{A Model of Two Interacting Carroll Particles}

In this section we will extend the analysis of the previous section and consider a model of two Carroll particles interacting through a potential V that depends on the
relative variables of the particles.
In order to construct the model we will first consider in subsection 3.1 a relativistic model of two interacting particles. Next, in subsection 3.2, we will consider the Carroll limit of this relativistic model. We will show that, unlike the single particle case, there is non-trivial dynamics in the two body system. It is sufficient to consider
the two relativistic particles in flat space-time  since, as we have seen in the single  particle case, the Carroll limit eliminates any reference to the
curvature of the spacetime we start with. Finally, in subsection 3.3 we will investigate the symmetries of the interacting Carroll model.

\subsection{A Relativistic Two Particle Model}

 We consider the interacting two relativistic particle model  of \cite{Todorov:1976pt,Komar:1978hc,DrozVincent:1978yk}. This model can be defined on  a phase space, where the coordinates and momenta are $x^{\mu}_i$, $p^{\mu}_i$, $(i = 1,2; \mu = 0,1,2,3)$.
The basic Poisson brackets are given by\,\footnote{Remember that the signature of $\eta$ is $(-,+++)$.}
 \be\{x_i^{\mu},\  p_j^{\nu}\} = \delta_{i,j}\eta_{\mu\nu}.
\ee
\color{black}
The model is defined in terms of two constraints $\phi_1$ and $\phi_2$
given by
\begin{eqnarray}\label{1bis}
\phi_1 = p_1^2 + m_1^2 + V\,, \hskip 2truecm
\phi_2 = p_2^2 + m_2^2 + V\,.
\nonumber\end{eqnarray}
We assume that the potential  V has the most general dependence on the variables allowed by the requirement that  the
following first class condition holds:
\begin{equation}\label{2}
\{\phi_1,\ \phi_2\} = 0\,.
\end{equation}
The constraints $\phi_1$ and $\phi_2$ are just a modification of the two mass shell constraints of two free particles
through the potential term.
This potential term breaks the two Poincar\'e invariances of the two free particles to a diagonal Poincar\'e
invariance\,\footnote{There is also a model where only  one combination of the mass shell constraints is first class and there is a further
tranversality constraint that is second class \cite{Kamimura:1977vi,Dominici:1978yu,Dominici:2013lba}.
This model has not a well defined Carroll limit and will not be considered here.}.

As shown, for instance in \cite{Rohrlich:1981ms}, the potential V can have a dependence on the set of scalars
that are formed from the variables $p_1^{\mu}$ and $p_2^{\mu}$, and the relative coordinate $r^{\mu} = x_1^{\mu} -
x_2^{\mu}$, transverse to the total momentum $P^{\mu} = p_1^{\mu} + p_2^{\mu}$.
This can be seen as follows. The first class condition (\ref{2}) can be written as
\begin{equation}\label{a}
p_1^{\mu}{\partial V\over \partial x_1^{\mu}} -
p_2^{\mu}{\partial V\over \partial x_2^{\mu}} = 0.
\end{equation}
We must add to this the requirement of translation invariance, that is
\begin{equation}\label{b}
{\partial V\over \partial x_1^{\mu}} =
- {\partial V\over \partial x_2^{\mu}}\,,
\end{equation}
so that the first class condition becomes
\begin{equation}\label{c}
P^{\mu}{\partial V \over \partial r^{\mu}} = 0\,.
\end{equation}
This shows that indeed V must depend on the scalars formed from $p_1\,,p_2$ and the part of $r$
that is transverse  to the total momentum $P^\mu$.

The allowed  scalars are given by
\begin{eqnarray}\label{3}
&&s_1 = - r_{\bot}^2\,,\quad
s_2 = (r_{\bot},\ p_1)\,,\quad
s_3 = (r_{\bot},\ p_2) = - s_2\,,\nonumber\\[.1truecm]
&&s_4 = - p_1^2\,,\quad
s_5 = -  p_2^2\,,\quad
s_6 = -  (p_1,\ p_2)\,.
\end{eqnarray}
In these equations the transverse relative variable $r_{\bot}$ is defined by
\begin{equation}\label{4}
r^{\mu}_{\bot} =  r^{\mu} - {(P\cdot r)\over P^2}P^{\mu},
\end{equation}
where $\vec {r} = \vec{x}_1- \vec{x}_2$ and $\vec{P} = \vec{p}_1 + \vec{p}_2$.

This finishes our discussion of the relativistic two body model. In the next subsection we will take the Carroll
limit of this model.

\subsection{The Carroll Limit of the Two Body Relativistic Model}

The Carroll limit of the relativistic two body model defined in the previous subsection  is defined by rescaling the canonical variables with a parameter $\omega$ as follows:
\begin{eqnarray}\label{13}\nonumber
&&p_1^{0} = \omega E_1,\quad p_2^{0} = \omega E_2,\quad
x_1^{0} = {1\over \omega} t_1\,,\\[.1truecm]
&& x_2^{0} = {1\over \omega} t_2,\quad
m_1 = \omega M_1,\quad m_2 = \omega M_2
\nonumber\end{eqnarray}
and taking the limit $\omega \rightarrow \infty$.

%
The Carroll symmetries of the model we are looking for should be given by
\begin{eqnarray}\label{8}
t_i^{\prime} &=& t_i + \vec{\beta}\cdot\vec{x}_i + a_t, \quad
\vec{x}_i^{\prime} = \vec{x}_i + \vec{a},\nonumber\\[.1truecm]
E^{\prime}_i &=& E_i,\quad
\vec{p}^{\prime}_i = \vec{p}_i + \vec{\beta}E_i, \quad (i = 1,2)\,.
\end{eqnarray}
These transformations are the Carroll limit of the diagonal Poincar\'e symmetries of the two particles.

In order to derive the dependence of the Carroll potential on the coordinates
we should isolate in the asymptotic expansion of the scalars those terms that scale like $\omega^2$:
\begin{eqnarray}\label{15}
s_4 &=& \omega^2 E_1^2 - (\vec{p}_1)^2 + O(\omega^{-2}),\quad\quad\quad
s_5 = \omega^2 E_2^2 - (\vec{p}_2)^2 + O(\omega^{-2})\,,\nonumber\\[.2truecm]
s_6 &=& \omega^2 E_1 E_2 - \vec{p}_1\cdot\vec{p}_2 + O(\omega^{-2})\,,\quad
s_4 s_5 - s_6^2 = - \omega^2(E_1\vec{p}_2 - E_2\vec{p}_1)^2+\cdots\,.
\end{eqnarray}
Note that in the last equation  the square of the
Carroll boost invariant vector
\be
\vec q \equiv E_1\vec p_2-E_2\vec p_1
\ee
occurs.

We conclude that  the potential could depend on $q^2\,, E_1$ and $E_2$, i.e., $V=V(\vec q^2, E_1,E_2)$.
Furthermore, since $\vec r=\vec x_1-\vec x_2$ is also
a Carroll boost invariant vector we could have the more general potential
$V=V(\vec{r}^2,\vec{q}^2,\vec{r}\cdot\vec{q}, E_1, E_2)$.
Therefore, the two first class Carroll invariant mass shell constraints $\phi_i\, (i=1,2)$ which satisfy  the constraint $\{\phi_1,\phi_2\}=0$ and  define the  model are given by 

\begin{equation}\label{10}
\phi_i = E_i^2 -M^2_i - V(\vec{r}^2,\vec{q}^2,\vec{r}\cdot\vec{q}, E_1, E_2)\,.
\end{equation}


Note that, due to the dependence of the potential on $(\vec{r})^2$ the two particle model is sensitive to the curvature of the transverse space. We consider here a flat transverse space.
This leads to the following  canonical action for the model
\begin{eqnarray}\label{9}
S &=& \int  d\tau\ \mathcal{L} = \int  d\tau\ \Big[
 E_1 \dot{t}_1 + E_2 \dot{t}_2
- \dot{\vec{x}}_1\cdot\vec{p}_1 - \dot{\vec{x}}_2\cdot\vec{p}_2
+\lambda_1\phi_1+\lambda_2\phi_2\Big]\,.
\end{eqnarray}
We observe  that the infinite-dimensional symmetries of the single particle are absent in the two-particle model
due to the non-invariance of the relative coordinates $\vec r$ and $\vec q$.

Having defined the interacting Carroll model, we are going to investigate  its dynamics in the next section.

\section{ The Dynamics of  Two Carroll Particles}

The equations of motion derived from the canonical action \eqref{9} are given by
\begin{eqnarray}\label{17}\nonumber
\dot{\vec{x}}_1 &=& \{\vec{x}_1,\ -V\}(e_1 + e_2)\,,\hskip 3truecm
\dot{\vec{x}}_2 = \{\vec{x}_2,\ -V\}(e_1 + e_2)\,,\nonumber\\[.1truecm]
\dot{t}_1 &=& - 2 E_1e_1 + \{t_1,\ - V\}(e_1 + e_2)\,,\hskip 1.3truecm
\dot{t}_2 = - 2 E_2 e_2 + \{t_2,\ - V\}(e_1 + e_2)\,,\nonumber\\[.1truecm]
\dot{\vec{p}}_1 &=& \{\vec{p}_1,\ - V\}(e_1 + e_2)\,,\hskip 3truecm
\dot{\vec{p}}_2 = \{\vec{p}_2,\ - V\}(e_1 + e_2)\,,\nonumber\\[.1truecm]
\dot{E}_1 &=& \{E_1,\ - V\}(e_1 + e_2)\,,\hskip 3truecm
\dot{E}_2 = \{E_2,\ - V\}(e_1 + e_2)\,.
\end{eqnarray}

We wish to investigate whether in general two interacting Carroll particles can move, i.e.~can have non-trivial dynamics.  In order to find out
we  choose as an example a simple potential that only depends on the ``relative momenta" $\vec q = E_1\vec{p}_2 - E_2\vec{p}_1$, i.e., we consider
a potential of the form  $V = V(\vec{q}^2)$.
In this special case the equations of motion of $\vec{x}_i$ read
\begin{eqnarray}\label{q8}
\dot{\vec{x}}_1 &=& - {4\over D} V^{\prime}E_2\vec{q}
(E_2 \dot{t}_1 + E_1 \dot{t}_2)\,,\hskip 1.5truecm
\dot{\vec{x}}_2 = + {4\over D} V^{\prime}E_1\vec{q}
(E_2 \dot{t}_1 + E_1 \dot{t}_2)\,,
\end{eqnarray}
where we have used the equations of motion \eqref{17} to write the Einbein variables in terms of $\dot{t}_i$ and
a new variable $D$ which is defined  by
\begin{equation}\label{q6}
D = 4\big[E_1 E_2 + E_1 V^{\prime}(\vec{p}_1\cdot\vec{q}) -
E_2 V^{\prime}(\vec{p}_2\cdot\vec{q}) \big]\,.
\end{equation}

We fix the two gauge symmetries generated by the two first class constraints $\phi_1$ and $\phi_2$ by imposing
 the gauge-fixing conditions $t_1 = t_2 = t$. Substituting these  conditions back into the equations of motion we obtain
\begin{eqnarray}\label{q9}
&&{d\vec{x}_1\over dt} =  - E_2{4(E_1+E_2)\over D}V^{\prime}\vec{q}\,,\hskip 1.5truecm
{d\vec{x}_2\over dt} = + E_1{4(E_1+E_2)\over D}V^{\prime}\vec{q}\,.
\end{eqnarray}
From this we derive that
\begin{equation}\label{q10}
{d \over dt}(E_1\vec{x}_1 + E_2\vec{x}_2) = 0.
\end{equation}
In this equation  we still need  to express $E_1$ and $E_2$ using the mass shells constraints \eqref{10}.
 For our choice of potential we can express $E_1$ and $E_2$ in terms of the momenta $p_1, p_2$ and the masses $M_1,M_2$.
 We have four sheets of solutions.
Introducing a small parameter
$\alpha$
in the potential we can write
\be
E_i=M_i+O_i(\alpha)\,.
\ee
We find that to lowest order in $\alpha$  the velocity of the center of mass is conserved, i.e.,
\be\label{move1}
M_1 {d\vec{x}_1\over dt} + M_2 {d\vec{x}_2\over dt}= {\rm constant}\,.
\ee
This implies non-trivial dynamics for the separate particles.

Our final conclusion is that, in contrast to a single Carroll particle, interacting Carroll particles can have non-trivial dynamics!

\section{Coupling the Carroll Particle to Gauge Fields}

In this section we will consider the gauging of the Carroll algebra and consider the coupling of the Carroll particle to the gauge fields corresponding to the Carroll algebra. In the first subsection we will introduce the Carroll algebra and compare it with the well-known Galilei algebra. It is known that a gauging of the Galilei algebra,
 or more precisely its centrally extended version, the Bargmann algebra, leads to a description of Newton-Cartan gravity. In the second subsection we will investigate what happens when one applies the same gauging procedure to the (non-centrally extended) Carroll algebra. Finally, in the last subsection we will consider the coupling of the Carroll particle to the gauge fields of the Carroll algebra.

\subsection{Comparing the  Galilei and Carroll Algebras}

The Galilei and Carroll algebras can be viewed as different contractions of the Poincar\'e algebra. We therefore start by considering the  Poincar\'e algebra in $D$ spacetime dimensions
\begin{eqnarray}
[M_{BC}, P_A] = -2\eta_{A[B}P_{C]}\,,\hskip 1.5truecm [M_{CD}, M_{EF}] = 4
\eta_{[C[E}\,M_{F]D]}\,,
\end{eqnarray}
where the indices $A,B,\dots = 0,1,\dots ,D-1$ are flat Lorentz indices.
Writing $A=(0,a)\,, a=1,2,\dots ,D-1$, the usual non-relativistic limit of the Poincar\'e algebra is defined by means of the following contraction:
\begin{equation}\label{Gcontraction}
\text{Galilei contraction:}\hskip 1truecm  P_0=\frac{1}{\omega}\,H\,,\hskip 1.5truecm M_{a0} = \omega\, G_a\,, \hskip 1truecm \omega\rightarrow \infty\,,
\end{equation}
which leads to the $D$-dimensional Galilei algebra:
\vskip .2truecm
\begin{center}
{\bf Galilei Algebra}
\end{center}
\vskip -.9truecm
\begin{eqnarray}\label{Galgebra}
&&[J_{ab}, P_c] = -2\delta_{c[a}P_{b]}\,,\hskip 1.5truecm [J_{ab}, G_c] = -2\delta_{c[a}G_{b]}\,,\nonumber\\[.2truecm]
&&[J_{cd}, J_{ef}] = 4
\eta_{[c[e}\,J_{f]d]}\,,\hskip 1.3truecm  [G_a, H] =-P_a\,.
\end{eqnarray}
We have renamed $M_{ab}= J_{ab}$. Here $(H, P_a, J_{ab}, G_a)$ are the generators of time translations, space translations, boosts and spatial rotations, respectively.

The Galilei transformations corresponding to the algebra \eqref{Galgebra} acting on spacetime
coordinates $x^\mu=(t,x^i), i=1,2,\dots ,D-1, $ are given by
\begin{equation}\label{Galtrans}
\delta t = -\zeta\,,\hskip 1.5truecm \delta x^i = \lambda^i{}_j x^j -v^i t -a^i\,.
\end{equation}
Here $(\zeta\,, \lambda^i{}_j\,, v^i\,, a^i)$ parametrize a (constant) time translation, space translation, spatial rotation and boost transformation, respectively.
A special feature of the Galilei algebra is that it admits a central extension. The centrally extended Galilei algebra contains an additional central charge generator $Z$ and is called the Bargmann algebra. The commutators of the Bargmann algebra are given by  those of the Galilei algebra, see eq.~\eqref{Galgebra}, together with the following commutator containing $Z$:
\begin{equation}\label{interchange}
[G_a, P_b] = -\delta_{ab}\,Z\,.
\end{equation}
It turns out that this central extension is indispensable in order to show that Newton-Cartan gravity follows from the gauging of an algebra.

There exists another less well-known contraction of the Poincar\'e algebra which corresponds to taking the ultra-relativistic limit. This so-called Carroll contraction
is given by \cite{Levy-Leblond}:
\begin{equation}\label{Ccontraction}
\text{Carroll contraction:}\hskip 1truecm  P_0=\omega\,H\,,\hskip 1.5truecm M_{a0} = \omega\, G_a\,, \hskip 1truecm \omega\rightarrow \infty\,.
\end{equation}
 This contraction leads to the
$D$-dimensional Carroll algebra \cite{Levy-Leblond}:
\vskip .2truecm
\begin{center}
{\bf Carroll Algebra}
\end{center}
\vskip -.9truecm
\begin{eqnarray}\label{Calgebra}
&&[J_{ab}, P_c] = -2\delta_{c[a}P_{b]}\,,\hskip 1.5truecm [J_{ab}, G_c] = -2\delta_{c[a}G_{b]}\,,\nonumber\\[.2truecm]
&&[J_{cd}, J_{ef}] = 4
\eta_{[c[e}\,J_{f]d]}\,,\hskip 1.3truecm  [G_a,P_b] = -\delta_{ab}H\,.
\end{eqnarray}
In contrast to the Galilei algebra, the Carroll algebra does not allow for a central extension\,\footnote{An
 exception is the 3D Carroll algebra which does allow a central extension of the form $[G_a,G_b]=\epsilon_{ab} {\tilde Z}$. Since we wish to consider the generic situation, valid for $D$ dimensions, we will not consider this central extension any further.}. The Carroll transformations corresponding to the algebra \eqref{Calgebra} acting on spacetime
coordinates $x^\mu=(t,x^i), i=1,2,\dots,D-1, $ are given by
\begin{equation}\label{Carrolltrans}
\delta t = -\zeta - v^i x^i\,,\hskip 1.5truecm \delta x^i = \lambda^i{}_j x^j  -a^i\,.
\end{equation}
Here, like in the Galilei case,  $(\zeta\,, \lambda^i{}_j\,, v^i\,, a^i)$ parametrize a (constant) time translation, space translation, spatial rotation and boost transformation, respectively.

Below we derive the gauge transformations of the Carroll gauge fields by gauging the Carroll algebra thereby stressing the common features as well as the differences with the gauging of the centrally extended Galilei algebra, i.e.~the Bargmann algebra.

\subsection{Gauging the Carroll Algebra}

We consider the gauging of the $D$-dimensional Carroll algebra \eqref{Calgebra} following the same gauging procedure that in the case of the  Bargmann algebra leads to Newton-Cartan gravity \cite{Andringa:2010it,Andringa:2013zja}
and see how far we can get.
As a first step we introduce for each generator of the Carroll algebra
a gauge field, a local parameter parametrizing the corresponding symmetry and the gauge-covariant curvatures, see Table \ref{Carrolltable}.

{\small
\begin{table}[t]
\begin{center}
\begin{tabular}{|c|c|c|c|c|}
\hline
symmetry&generators& gauge field&parameters&curvatures\\[.1truecm]
\hline\rule[-1mm]{0mm}{6mm}
temporal translations&$H$&$\tau_\mu$&$\zeta(x^\nu)$&$R_{\mu\nu}(H)$\\[.1truecm]
spatial translations&$P_a$&$e_\mu{}^a$&$\zeta^a(x^\nu)$&$R_{\mu\nu}{}^a(P)$\\[.1truecm]
boosts&$G_a$&$\omega_\mu{}^{a}$&$\lambda^a(x^\nu)$&$R_{\mu\nu}{}^a(G)$\\[.1truecm]
spatial rotations&$J_{ab}$&$\omega_\mu{}^{ab}$&$\lambda^{ab}(x^\nu)$&$R_{\mu\nu}{}^{ab}(J)$\\[.1truecm]
\hline
\end{tabular}
\caption{This table indicates the generators of the Carroll algebra and the  gauge fields, local parameters and curvatures that are associated to each of these generators.}\label{Carrolltable}
\end{center}
 \end{table}
 }
According to the Carroll algebra \eqref{Calgebra} the gauge fields transform
as follows:\footnote{All parameters depend on the coordinates $x^\mu$, even when not explicitly indicated.}
\begin{align} \label{bossymm1}
\delta \tau_\mu &=  \lambda^a e_{\mu a} +\partial_\mu\zeta-\zeta^a\omega_\mu{}^{a}\,, \nonumber \\[.1truecm]
\delta e_\mu{}^a &= (D_\mu\zeta)^a +  \lambda^a{}_b e_\mu{}^b\,, \nonumber \\[.1truecm]
\delta \omega_\mu{}^{ab} &=  \partial_\mu \lambda^{ab}\,, \\[.1truecm]
\delta \omega_\mu{}^{a} &= (D_\mu\lambda)^a +  \lambda^{a}{}_b \omega_{\mu }{}^b
\nonumber \,,
\end{align}
where $D_\mu$ is the covariant derivative with respect to spatial rotations.
The following curvatures transform covariantly under these transformations:
\begin{align}\label{curvatures}
R_{\mu \nu}(H) &=  2 \partial_{[\mu} \tau_{\nu]} - 2\omega_{[\mu}{}^a e_{\nu] a}\,, \nonumber
\\[.1truecm]
R_{\mu \nu}{}^a(P) &=  2 \partial_{[\mu} e_{\nu]}{}^a - 2
\omega_{[\mu}{}^{ab} e_{\nu]b} \,, \nonumber \\[.1truecm]
R_{\mu\nu}{}^a(G) &= 2\partial_{[\mu}\omega_{\nu]}{}^{a} - 2
\omega_{[\mu}{}^{ab} \omega_{\nu]b}\,,
\\[.1truecm]
R_{\mu\nu}{}^{ab} (J) &=
2\partial_{[\mu}\omega_{\nu]}{}^{ab}  \,. \nonumber
\end{align}

Our first task is now to impose  conventional constraints on the curvatures, like one does when gauging the Bargmann algebra \cite{Chamseddine:1976bf}.  To be precise, we impose constraints on the Carroll curvatures \eqref{curvatures} such that the temporal  and spatial translations, with parameters $\zeta$ and $\zeta^a$, get equivalent to the general coordinate transformations, with parameters $\xi^\mu$,  modulo
boosts, with parameters $\lambda^a = \xi^\mu\omega_\mu{}^a$  and spatial rotations, with parameters $\lambda^{ab} = \xi^\mu\omega_\mu{}^{ab}$. For this, we need the following two identities for those  gauge fields
that transform under $H$ and/or $P_a$-transformations.:
\begin{eqnarray}
\delta_{\text{g.c.t.}}(\xi^\nu)\tau_\mu &=& \Big[\delta_H(\x^\lambda\tau_\lambda) + \delta_{P}(\xi^\lambda e_\lambda{}^a) + \delta_{G}(\xi^\lambda\omega_\lambda{}^a)\Big]\tau_\mu+\xi^\lambda R_{\lambda\mu}(H)\,,\\[.1truecm]
\delta_{\text{g.c.t.}}(\xi^\nu)e_\mu{}^a &=& \Big[\delta_{P}(\xi^\lambda e_\lambda{}^a) +\delta_{J}(\xi^\lambda\omega_\lambda{}^{ab})\Big] e_\mu{}^a +\x^\lambda R_{\lambda\mu}{}^a(P)\,.
\end{eqnarray}
These identities show that, in order to equate a general coordinate transformation to an $H$- and $P_a$-transformation, modulo a boost and/or
a spatial rotation, we  need to impose the following set of conventional constraints:
\begin{equation}\label{conventional}
R_{\mu\nu}(H) = R_{\mu\nu}{}^a(P) =0\,.
\end{equation}
We furthermore deduce that the relation between the different parameters is given by
\begin{equation}
\zeta=\xi^\mu\tau_\mu\,,\hskip 2truecm  \zeta^a=\xi^\mu e_\mu{}^a\,.
 \end{equation}
Introducing the projective inverses $\tau^\mu$ and $e^\mu{}_a$ of $\tau_\mu$ and $e_\mu{}^a$, respectively, as
follows:
\begin{alignat}{2} \label{inverses}
e_\mu{}^a e^\mu{}_b &= \delta^a_b \,, &\qquad \tau^\mu \tau_\mu = 1 \,, \nonumber \\[.2truecm]
\tau^\mu e_\mu{}^a &= 0 \,, & \tau_\mu e^\mu{}_a = 0 \,, \\[.2truecm]
e_\mu{}^a e^\nu{}_a &= \delta^\nu_\mu - \tau_\mu \tau^\nu \,. & \nonumber
\end{alignat}
 we derive that the inverse relation between $\zeta,\zeta^a$ and $\xi^\mu$ is given by
\begin{equation}
 \xi^\mu = \tau^\mu\zeta + e^\mu{}_a\zeta^a\,.
 \end{equation}

The gauge fields $\tau_\mu$ and $e_\mu{}^a$ can now be interpreted as the temporal and spatial Vielbeine.
The transformations of these Vielbeine together with their  projective inverse fields under boosts and spatial rotations are given by
\begin{eqnarray}
\delta\tau_\mu &=& \lambda^a e_{\mu a}\,, \hskip 2truecm \delta e_\mu{}^a = \lambda^a{}_b e_\mu{}^b\,,\\[.1truecm]
\delta\tau^\mu &=& 0\,,\hskip 2.8truecm \delta e^\mu{}_a = \lambda_a{}^b e^\mu{}_b - \tau^\mu\lambda_a\,,
\end{eqnarray}
while under general coordinate transformations they transform as covariant ($\tau_\mu$ and $e_\mu{}^a$) and contra-variant ($\tau^\mu$ and $e^\mu{}_a$) vectors.
From now on we will work solely with the general coordinate transformations and not consider the temporal and spatial translations anymore.

We observe that the projective invertability relations \eqref{inverses} are invariant under the non-trivial
boost transformations of $\tau_\mu$ and $e^\mu{}_a$. Like in the Bargmann case, this corresponds to the ambiguity of defining the inverse of a singular matrix. Therefore, in the Carroll case the only fields that are un-ambigiously defined are
\begin{equation}
\text{Carroll:}\hskip 1truecm \{\tau^\mu\,, e_\mu{}^a\}
\end{equation}
These Carroll fields are invariant under boosts and transform  in the standard way under spatial rotations and general coordinate transformations. Note that there is no central charge gauge field.

Unlike in the Bargmann case, the conventional constraints \eqref{conventional} are not sufficient to solve for the
boost gauge fields $\omega_\mu{}^a$ and the gauge field of spatial rotations $\omega_\mu{}^{ab}$. The reason for this difference is that the Bargmann algebra leads to an additional central charge gauge field whose curvature may be set to zero. This particular conventional constraint plays a crucial role in solving for the spin-connection fields in the Bargmann case.
In the Carroll case, the
boost gauge field $\omega_\mu{}^a$ only occurs in the first constraint in \eqref{conventional} and this constraint is
invariant under the following shift symmetries:
\begin{equation}
\omega_\mu{}^a \ \rightarrow \omega_\mu{}^a + e_{\mu b} X^{(ab)}\,,
\end{equation}
with  $X^{(ab)}$ an arbitrary symmetric tensor. This shows that $\omega_\mu{}^a$ can only be solved modulo this ambiguity.


Another difference with the Galilei case is that a foliation-defining constraint, like the
constraint $\partial_{[\mu}\tau_{\nu]G}=0$ that we have in the Bargmann case, is absent. The reason for this is that in the Carroll case such a constraint is not invariant under
boost transformations. Instead, it is the inverse temporal Vielbein $(\tau^\mu)_C$ that is invariant under boost transformations and hence has an invariant meaning. This is in line with the duality relation between
 the Galilei and Carroll cases discussed in \cite{Houlrik:2010rd,Duval:2014uoa}
\be
(\tau_\mu dx^\mu)_G \  \  \ \longleftrightarrow \  \  \ (\tau^\mu\frac{\partial}{\partial x^\mu})_C\,.
\ee

\subsection{Coupling the Carroll Particle to  Gauge Fields}

Having established in the previous subsection the transformation  rules of the Carroll gauge fields, let us now couple a Carroll particle to these gauge fields. Since $\tau^\mu$ is invariant under boosts it is natural to use this gauge field
when coupling to the Carroll particle.  The covariantization of the
action \eqref{0} which is invariant under the sigma model symmetries is
\be\label{actioncurved}
S=\int d\tau\,\Big[p_\mu\dot{ x^\mu}-\frac e2\Big(\tau^\mu(t,\vec x)\tau^\nu(t,\vec x) p_\mu p_\nu-M^2\Big)\Big]\,,
\ee
where $p_0=-E$.
The equations of motion that follow from this action are given by
\be\label{move2}
\dot x^\mu=e\tau^\mu\tau^\nu p_\nu,\quad \dot p_\mu=-e(\partial_\mu\t^\rho)\tau^\sigma p_{\sigma} p_{\rho}\,.
\ee
From these equations we deduce that the single Carroll particle in a non-trivial background specified by the inverse
Vielbein $\tau^\mu$ has non-trivial dynamics.

It is not difficult  to compute the Killing equations to find the Noether symmetries of the action \eqref{actioncurved}.
We take as the generator of these transformations
\be
G=\xi^\mu (t,\vec x)p_\mu\,.
\ee
 For a flat transverse space, the condition of being a symmetry, $\dot G=0$, implies
\be
{\cal L}_\xi\tau^\mu=0\,.
\ee
If curvature is turned on by hand this condition should be supplemented with the requirement that ${\cal L}_\xi g_{ij}=0$
\cite{Duval:2014uoa}.


\section{Discussion}

 The particle models we have considered in this paper are obtained by taking the Carroll limit of the relativistic particle. The resulting Carroll particle is by construction invariant under the Carroll symmetries.
 A particular feature of the Carroll limit is that it  wipes out all information about the curvature of the spacetime in which the original relativistic particle was moving in. The free Carroll particle has several noteworthy features.
 First of all, the free particle action is in fact invariant under infinite-dimensional symmetries,
 see eq.~\eqref{solKil}. Secondly,
 the mass-shell constraint allows for positive as well as negative energy solutions, like in the relativistic case.
 There exists also the limit to a massless Carroll particle.

Concerning the dynamics, the free Carroll particle has rather
uninteresting dynamics: it cannot move.
In this paper we investigated two situations in which this is no longer the case. We first showed that for a set of interacting
Carroll particles only the center of mass cannot move but that the separate particles can have non-trivial dynamics,
see eq.~\eqref{move1}. Next, we introduced non-trivial background fields by gauging the Carroll algebra and coupled these background fields to a single
Carroll particle. We showed that even a single Carroll particle, due to the non-trivial
 background fields, can move, see eq.~\eqref{move2}.

When gauging the Carroll algebra, we were not able to find a set of conventional constraints that enables one to solve for the connection fields corresponding to spatial rotations and boosts. This is in contrast to what happens when gauging the Bargmann algebra where the connection fields can be solved and where the gauging procedure  leads to a
description of Newton-Cartan gravity. A crucial role is here played by the central charge generator
which can be added to the Galilei algebra but not to the Carroll algebra. It would be interesting to see in which sense a definition of `Carroll gravity' can be given consistent with the duality between Galilei and Carroll gravity proposed in
\cite{Houlrik:2010rd,Duval:2014uoa}.

One direction where Caroll symmetries and the non-trivial particle dynamics could have applications is in studies of the
gauge/gravity duality where BMS symmetries, which are closely related to Carroll symmetries \cite{Duval:2014uva,Duval:2014uoa}, emerge at the boundary. It would be interesting to investigate these relationships further and, in particular, to consider
generalizations of the Carroll contraction that correspond to the ultra-relativistic limit of a world probed by extended objects such as strings and branes.


\section*{Acknowledgements}
E.B. and J.G. thank Roberto Casalbuoni, Gary Gibbons, Peter Horvathy and Jan Rosseel for useful discussions. J.G. acknowledges the hospitality at the Department of Theoretical Physics of the University of Groningen where this work was done.
J.G. also acknowledges partial financial support from  the Dutch research organisation FOM and from
FPA 2010-20807, 2009 SGR502, CPAN, Consolider CSD 2007-0042.

\begin{appendix}

\section{An Alternative Derivation of the Carroll Particle Action}

In this appendix we give an alternative derivation of  the Carroll particle action \eqref{1} using the non-linear realization method  \cite{Coleman}. Our starting point is the $D$-dimensional Carroll alrebra given in \eqref{Calgebra}.
We consider the coset $\frac{{\rm Carroll}}{{\rm Rotations}}$ \cite{old}, that we locally parametrize as
\be
g=e^{-tH}e^{\vec x\vec P}e^{\vec v\vec G}\,,
\ee
where $ t, \vec x$ are the Goldstone bosons associated to space-time translations and $\vec v$ are the Goldstone bosons  associated to the broken boosts.

The Maurer-Cartan form $\Omega$ is given by
\bea
\Omega=g^{-1}dg&=&H L^H+L^{\vec P}\vec P+L^{\vec G}\vec G
\\
&=& (-dt+\vec v d\vec x)H+d\vec x \vec P+d\vec v\vec G\,.
\eea
The action for the particle with lowest order in derivatives is obtained by taking the pullback of the rotation invariant form $L^H$ to the
world-line of the particle:
\be
S=M\int d\tau( {L^H})^*=\int d\tau(-M\dot t+M\vec v\dot{\vec x})\,.
\ee
The momentum of the Carroll particle is therefore $\vec p=M\vec v.$ The Goldstone bosons of the broken boosts are always related to the momentum of the particle. The action obtained from the non-linear realization
in general is a phase space action,
see for example  \cite{Gomis:2012ki}.

Finally, the transformations that leave invariant the Maurer-Cartan form are  the Carroll symmetries
\bea\label{Ctransf1}
&&t'=t+\vec\beta\cdot\vec x+a_t\,,\hskip 2truecm
\vec x'=\vec x+\vec a\,,
\nn\\
&&\vec v'=\vec v+ \vec\beta\,. \hskip 2.9truecm
\eea

\end{appendix}


\begin{thebibliography}{99}

\bibitem{Bacry:1968zf}
  H.~Bacry and J.~Levy-Leblond,
  ``Possible kinematics,''
  J.\ Math.\ Phys.\  {\bf 9} (1968) 1605.


\bibitem{dubouis}
J.~Derome and J. G.~Dubouis,``Hooke's symmetries and nonrelativistic cosmological model kinematics,"
Nuovo Cimento B {\bf 9} (1972) 351.

\bibitem{Levy-Leblond}
J.M.~L\'evy-Leblond, ``Une nouvelle limite non-relativiste du group de Poincar\'e'',
Ann.~Inst.~H.~Poincar\'e {\bf 3} (1965) 1.


\bibitem{Henneaux:1979vn}
  M.~Henneaux,
  ``Geometry of Zero Signature Space-times,''
  Bull.\ Soc.\ Math.\ Belg.\  {\bf 31} (1979) 47;
  M.~Henneaux, M.~Pilati and C.~Teitelboim,
  `Explicit Solution for the Zero Signature (Strong Coupling) Limit of the Propagation Amplitude in Quantum Gravity,''
  Phys.\ Lett.\ B {\bf 110} (1982) 123.

\bibitem{Dautcourt:1997hb}
  G.~Dautcourt,
  ``On the ultrarelativistic limit of general relativity,''
  Acta Phys.\ Polon.\ B {\bf 29} (1998) 1047
  [gr-qc/9801093].


\bibitem{Duval:2014uva}
  C.~Duval, G.~W.~Gibbons and P.~A.~Horvathy,
  ``Conformal Carroll groups and BMS symmetry,''
  arXiv:1402.5894 [gr-qc].



\bibitem{Bondi:1962px}
  H.~Bondi, M.~G.~J.~van der Burg and A.~W.~K.~Metzner,
  ``Gravitational waves in general relativity. 7. Waves from axisymmetric isolated systems,''
  Proc.\ Roy.\ Soc.\ Lond.\ A {\bf 269} (1962) 21;
  R.~Sachs,
  ``Asymptotic symmetries in gravitational theory,''
  Phys.\ Rev.\  {\bf 128} (1962) 2851.


\bibitem{Barnich:2009se}
  G.~Barnich and C.~Troessaert,
  ``Symmetries of asymptotically flat 4 dimensional spacetimes at null infinity revisited,''
  Phys.\ Rev.\ Lett.\  {\bf 105} (2010) 111103
  [arXiv:0909.2617 [gr-qc]].

\bibitem{Bagchi:2012cy}
  A.~Bagchi and R.~Fareghbal,
  ``BMS/GCA Redux: Towards Flatspace Holography from Non-Relativistic Symmetries,''
  JHEP {\bf 1210} (2012) 092
  [arXiv:1203.5795 [hep-th]].


\bibitem{old}
J. Gomis and F. Passerini, unpublished notes (2005).


\bibitem{Ngendakumana:2013hza}
  A.~Ngendakumana, J.~Nzotungicimpaye and L.~Todjihounde,
  ``Group theoretical construction of planar Noncommutative Phase Spaces,''
  J.\ Math.\ Phys.\  {\bf 55} (2014) 013508
  [arXiv:1308.3065 [math-ph]].

\bibitem{Duval:2014uoa}
  C.~Duval, G.~W.~Gibbons, P.~A.~Horvathy and P.~M.~Zhang,
``Carroll versus Newton and Galilei: two dual non-Einsteinian concepts of time,''  arXiv:1402.0657 [gr-qc].  



\bibitem{Coleman}
  S.~R.~Coleman, J.~Wess and B.~Zumino,
  ``Structure of phenomenological Lagrangians. 1',
  Phys.\ Rev.\  {\bf 177} (1969) 2239;
  \newline
  C.~G.~.~Callan, S.~R.~Coleman, J.~Wess and B.~Zumino,
  {``Structure of phenomenological Lagrangians. 2''},
  Phys.\ Rev.\  {\bf 177} (1969) 2247.


\bibitem{Gomis:2000bd}
  J.~Gomis and H.~Ooguri,
  ``Nonrelativistic closed string theory,''
  J.\ Math.\ Phys.\  {\bf 42} (2001) 3127
  [hep-th/0009181].

\bibitem{Gomis:2005pg}
  J.~Gomis, J.~Gomis and K.~Kamimura,
  ``Non-relativistic superstrings: A New soluble sector of AdS(5) x S**5,''
  JHEP {\bf 0512} (2005) 024
  [hep-th/0507036].








\bibitem{Duval:2014lpa}
  C.~Duval, G.~W.~Gibbons and P.~A.~Horvathy,
  ``Conformal Carroll groups,''
  arXiv:1403.4213 [hep-th].

\bibitem{Todorov:1976pt}
  I.~T.~Todorov,
  ``Dynamics of Relativistic Point Particles as a Problem with Constraints,''
  JINR-E2-10125.

\bibitem{Komar:1978hc}
  A.~Komar,
  ``Constraint Formalism of Classical Mechanics,''
  Phys.\ Rev.\ D {\bf 18} (1978) 1881.


\bibitem{DrozVincent:1978yk}
  P.~Droz-Vincent,
  ``Action-at-a-distance and Relativistic Wave Equations for Spinless Quarks,''
  Phys.\ Rev.\ D {\bf 19} (1979) 702.

\bibitem{Kamimura:1977vi}
  K.~Kamimura and T.~Shimizu,
  ``Relativistic Lagrangian for Multilocal Model,''
  Prog.\ Theor.\ Phys.\  {\bf 58}, 383 (1977).

\bibitem{Dominici:1978yu}
  D.~Dominici, J.~Gomis and G.~Longhi,
  ``A Lagrangian for Two Interacting Relativistic Particles: Canonical Formulation,''
  Nuovo Cim.\ A {\bf 48} (1978) 257.


\bibitem{Dominici:2013lba}
  D.~Dominici, J.~Gomis, K.~Kamimura and G.~Longhi,
  ``Dynamical sectors of a relativistic two particle model,''
  Phys.\ Rev.\ D {\bf 89} (2014) 045001
  [arXiv:1312.1000 [hep-th]].







\bibitem{Rohrlich:1981ms}
  F.~Rohrlich,
  In *Barcelona 1981, Proceedings, Relativistic Action At A Distance: Classical and Quantum Aspects*, 190-212

\bibitem{Andringa:2010it}
  R.~Andringa, E.~Bergshoeff, S.~Panda and M.~de Roo,
  ``Newtonian Gravity and the Bargmann Algebra,''  Class.\ Quant.\ Grav.\  {\bf 28} (2011) 105011  [arXiv:1011.1145 [hep-th]].  


\bibitem{Andringa:2013zja}
  R.~Andringa, E.~A.~Bergshoeff, J.~Rosseel and E.~Sezgin,
  ``3D Newton�Cartan supergravity,''  Class.\ Quant.\ Grav.\  {\bf 30} (2013) 205005.  



\bibitem{Chamseddine:1976bf}
This approach was used in the context of supergravity in A.~H.~Chamseddine and  P.~C.~West,
  ``Supergravity as a Gauge Theory of Supersymmetry,''  Nucl.\ Phys.\ B {\bf 129} (1977) 39. For more literature, see the textbooks  T.~Ort\`in,
  ``Gravity and strings,''
  (Cambridge University Press (2004)),  D.~Z.~Freedman and A.~Van Proeyen,
  ``Supergravity,''
  (Cambridge University Press (2012)) and   P.~West,
  ``Introduction to strings and branes,''   (Cambridge University Press (2012)).


\bibitem{Houlrik:2010rd}
  J.~M.~Houlrik and G.~Rousseaux,
  ``'Nonrelativistic' kinematics: Particles or waves?,''  arXiv:1005.1762 [physics.gen-ph].  




\bibitem{Gomis:2012ki}
  J.~Gomis, K.~Kamimura and J.~M.~Pons,
  ``Non-linear Realizations, Goldstone bosons of broken Lorentz rotations and effective actions for p-branes,''
  Nucl.\ Phys.\ B {\bf 871} (2013) 420
  [arXiv:1205.1385 [hep-th]].











\end{thebibliography}
\end{document}